\documentclass{cjpsuppl}
\usepackage{epsfig}
\begin{document}
\title{Search for TeV-scale bosons in the dimuon channel at the LHC.}
\authori{I.~Golutvin, P.~Moissenz, V.~Palichik, M.~Savina, \underline{S.~Shmatov}}
\addressi{Joint Institute for Nuclear Research, Dubna, Russia}
\authorii{}    \addressii{}
\authoriii{}   \addressiii{}
\authoriv{}    \addressiv{}
\authorv{}     \addressv{}
\authorvi{}    \addressvi{}
\headtitle{Search for TeV-scale bosons in the dimuon channel with the CMS detector}
\headauthor{I.~Golutvin, P.~Moissenz, V.~Palichik, M.~Savina, S.~Shmatov}
\lastevenhead{I.~Golutvin {\it et al.}: Search for TeV-scale bosons in the \ldots}
\pacs{62.20}
\keywords{extra gauge boson, extra dimensions, Randall-Sundrum, KK-graviton,
high-$p_T$ muons}
\refnum{}
\daterec{29 September 2003;\\final version 31 December 2003}
\suppl{A}  \year{2003} \setcounter{page}{21}
\maketitle

\begin{abstract}
   Extended gauge models and the Randall-Sundrum model with
   extra-dimension predict the existence of TeV-mass resonances.
   The LHC potential for a five sigma level discovery was
   investigated as described in this document.
   Final states containing large invariant mass di-muons from
   $\mbox{Z}^{\prime}$ and the RS1 graviton were studied.
   The possibility of discriminating between different $\mbox{Z}^{\prime}$
   model by measuring the muon forward-backward asymmetry was
   investigated. The determination of the spin of the resonance is also discussed.
\end{abstract}

\setcounter{page}{2}

\section{Introduction}

The Standard Model (SM) has been tested by the experiments at LEP,
SLC and Tevatron with a high accuracy. 
In particular, the yield of lepton pairs produced mainly via
Drell-Yan processes, i.e. quark-antiquark annihilation by exchange
of photons or $\mbox{Z}$ bosons, is predicted by the SM with a per
mille precision. So far, the experimental data have shown no
significant deviation from the SM predictions for the Drell-Yan
continuum up to TeV-energy-scale. The high-order calculations of
lepton pair production cross section in the mass region of $0.1
\div 1$ $\mbox{TeV}/c^2$ are indeed in good agreement with D0 and
CDF data \cite{D0, CDF}.


At present, however, there are many theoretical attempts to extend
the bounds of the SM in order to incorporate the unification of
strong and electroweak interactions, the mass hierarchy and
CP-violation problems, the arbitrariness of flavor mixing and the
number of generations, etc. Supersymmetry is the most popular
theoretical extension to the SM. However, as an alternative, also
extended gauge models, based on symmetry groups wider than in the
SM, have been considered. This way leads to the various left-right
symmetric models, extended gauge theories including grand
unification theories, and models of composite gauge bosons
\cite{EGM, GUT_EGM}. In all these cases, new vector bosons,
neutral $\mbox{Z}^{\prime}$ and charged $W^{\prime}$, would appear
at a mass scale of the order of one $\mbox{TeV}/c^2$ what can be
observed at LHC.


One of the most attractive and exciting, although the most
complicated, goal of modern theoretical physics is to provide a
"unified" description of all forces known in nature and give an
adequate explanation of the creation and the evolution of the
Universe. The quite new paradigm of the gravity at TeV energies,
as given in the large or infinite extra dimension (LED) and brane
world scenarios, which propose a solution of gauge hierarchy
problem, have recently discussed \cite{BW}.

One of the possible LED scenarios is the Randall-Sundrum (RS)
approach \cite{RS1} based on the warp phenomenology of the
$\mbox{AdS}_5$ nonfactorizable geometry with the curvature $k$
$\sim$ $M_{\mbox{\small Pl}}$ $\sim$ $10^{19}$ $\mbox{GeV}/c^2$
and the metric

$$ds_2 = e^{-kr_c\phi} \eta_{\mu\nu}dx^{\mu}dx^{\nu} + r^2_cd\phi^2.$$
\noindent Here, $r_c$, is the compactification radius of the extra
dimension, $\eta_{\mu\nu}$ is the standard four-dimensional
Minkowski metric, $x^{\mu}$ are ordinary four-dimension
coordinates and $\phi$ is the extra dimension coordinate. The
distinctive feature of the RS phenomenology with two branes (one
with positive tension $\sigma$ at $\phi=0$ and the other with
negative tension $-\sigma$ at distance $r_c$), so-called RS1 model
\cite{RS1}, is the infinite tower of Kaluza-Klien graviton modes
appearing at the scale $\Lambda_{\pi} = M_{\mbox{\small Pl}}
e^{-kr_c\pi}$. It has a zero mode, $m_0 \sim e^{-kr_c}$,
describing the usual four-dimensional gravity and massive modes
with the mass splitting between them of order $\Delta m \sim k
e^{-kr_{c}}$. The exponential factor $e^{-kr_c\pi}$ removes the
hierarchy between the Plank and electroweak scales if $kr_c
\approx$ $11 \div 12$. The first Kaluza-Klein graviton mode
(called the RS1 graviton below), as well as $\mbox{Z}^{\prime}$,
is strongly coupled to ordinary particles.

In both the conceptions above, the width of the predicted
resonances is not fixed, it can vary widely depending on the model
parameters. It implies that these states can appear as individual
resonances or can affect the high-$p_{\mbox{\small T}}$ lepton
pairs continuum leading to an excess of Drell-Yan production.
Thus, the distinctive experimental signature for these processes
is a pair of well-isolated high-$p_{\mbox{\small T}}$ leptons with
opposite charges coming from the same vertex.

These measurements can be performed at the both LHC experiments,
ATLAS and CMS, which is expected to be able to trigger and
identify hard muons with a transverse momentum up to several TeV.
The ability of the LHC experiments to detect virtual RS1 graviton
in the muon mode was investigated as described in
Ref.~\cite{RS1_CMS}. In this paper, the analysis of the LHC
discovery limit of $\mbox{Z}^{\prime}$ and both virtual and real
RS1 gravitons in assumption of the CMS acceptance are presented.

\section{Extra gauge bosons}

\subsection{Signal and background simulations}

The signal simulation is done for the parton subprocess $\mbox{q}
\bar{\mbox{q}} \to \mbox{Z}^{\prime}$ in the QCD leading order
without high order corrections. To generate the
$\mbox{Z}^{\prime}$ boson and its decay to a muon pair as well as
the relevant background events, the PYTHIA 6.217 \cite{Pythia}
package with the CTEQ5L parton distribution function was used.
There are many possible non Standard Model scenarios which predict
the existence of heavy neutral $\mbox{Z}^{\prime}$ and/or charged
$\mbox{W}^{\prime}$ gauge bosons (reviews in \cite{EGM}). The
$\mbox{Z}^{\prime}$ models which were used for this analysis are
the following:

\begin{enumerate}

\item
   The Left-Right model (LR) \cite{LRM} based on the electroweak gauge group symmetry
   $\mbox{SU}(2)_L \times \mbox{SU}(2)_R \times \mbox{U}(1)_{B-L}$ (here, $B$ and $L$ are the baryon and
   lepton numbers) with default PYTHIA couplings which are the same for both left- and right-handed type
   of fermions and are set to the same values as in the SM, $g_L$ = $g_R$ = 0.64. The number of
   extra fermion generations is equal to three.

\item
   $\mbox{Z}^{\prime}_{\chi}$-,  $\mbox{Z}^{\prime}_{\eta}$-, and $\mbox{Z}^{\prime}_{\psi}$-models which naturally arise
   as a result of the sequential breaking of $\mbox{SO}(10)$ or $\mbox{E}_6$ group symmetry
   for Grand Unified Theories (GUT) \cite{GUT_EGM}:
   $\mbox{E}_6 \to \mbox{SO}(10) \times \mbox{U}(1)_{\psi} \to \mbox{SU}(5) \times \mbox{U}(1)_{\chi} \times
   \mbox{U}(1)_{\psi} \to \mbox{SM} \times \mbox{U}(1)_{\theta_6}$. The linear combination of the hypercharges of the two groups
   $\mbox{U}(1)_{\chi} \times \mbox{U}(1)_{\psi}$ gives the charge of the lightest $\mbox{Z}^{\prime}$
   at the symmetry-breaking energies
   $\mbox{Z}^{\prime} = \mbox{Z}^{\prime}_{\chi} \mbox{cos}(\theta_{E_6}) + \mbox{Z}^{\prime}_{\psi}
   \mbox{sin}(\theta_{E_6})$.
   The numerical values of the couplings for these models are taken from
   Ref.~\cite{Zp_coupl}

\item
   Also used for Monte Carlo studies is the "sequential" standard
   model (SSM) \cite{SSM} in which the heavy bosons ($\mbox{Z}^{\prime}$
   and $\mbox{W}^{\prime}$) are assumed to couple only to one fermion type
   (left) with the same parameters (couplings and the total width) as for ordinary
   $\mbox{Z}^0$ and $\mbox{W}^{\pm}$ in the Standard Model.

\end{enumerate}

\begin{figure}[ht]
\vspace{-0.5cm}
\begin{center}
\resizebox{6.2cm}{!}{\includegraphics{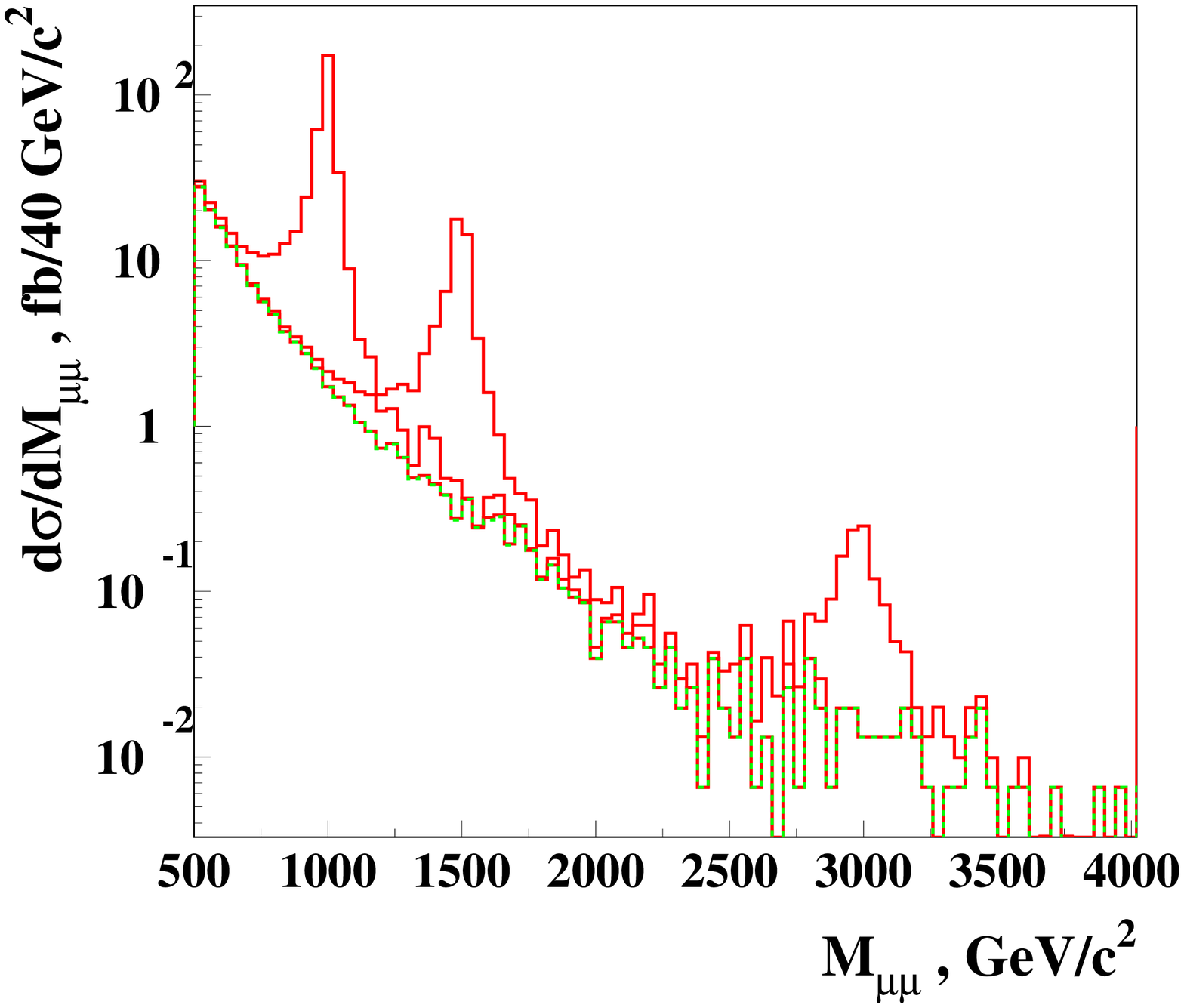}}
\resizebox{6.2cm}{!}{\includegraphics{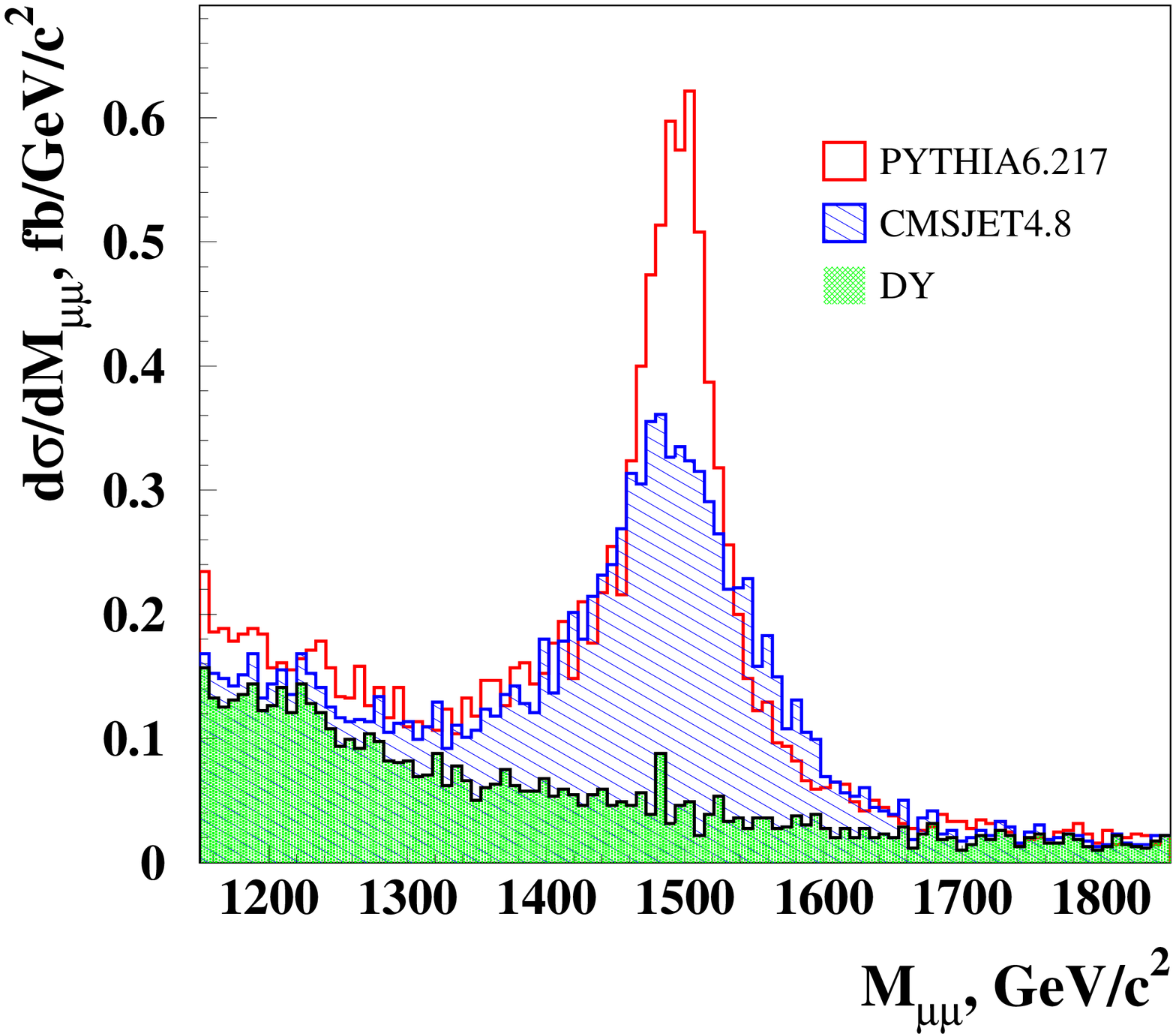}}
\vspace{-0.5cm} \caption{Production cross section of muon pairs as
a function of their invariant mass. Muins from
$\mbox{Z}^{\prime}_{\chi}$ decay over Drell-Yan continuum, as
generated by PYTHIA 6.217 (left), and after muon smearing (right).
The generated distribution is presented as the open histogram, the
detector response by line hatched one, and the Drell-Yan
background after muon smearing by full hatched one.}
\label{fig:crs_Z}
\end{center}
\end{figure}

The non-reducible background considered is the Drell-Yan processes
$\mbox{pp} \to \mbox{Z}/\gamma \to \mu^+ \mu^-$ which gives nearly
95 \% of the the Standard Model muon continuum. The contribution
from the other reaction (vector boson pair production
($\mbox{ZZ}$, $\mbox{WZ}$, $\mbox{WW}$), and $t \bar t$
production) is very small and is neglected in this study. In the
SM the expected number of di-muon events is not very large and the
$\mbox{Z}^{\prime}$ resonance peak exceeds the background by about
a factor ten (Fig.~\ref{fig:crs_Z}, left plot).


\subsection{$\mbox{Z}^{\prime}$ discovery limits}

\begin{figure}[ht]
\vspace{1.7cm}
\begin{center}
\resizebox{6.2cm}{!}{\includegraphics{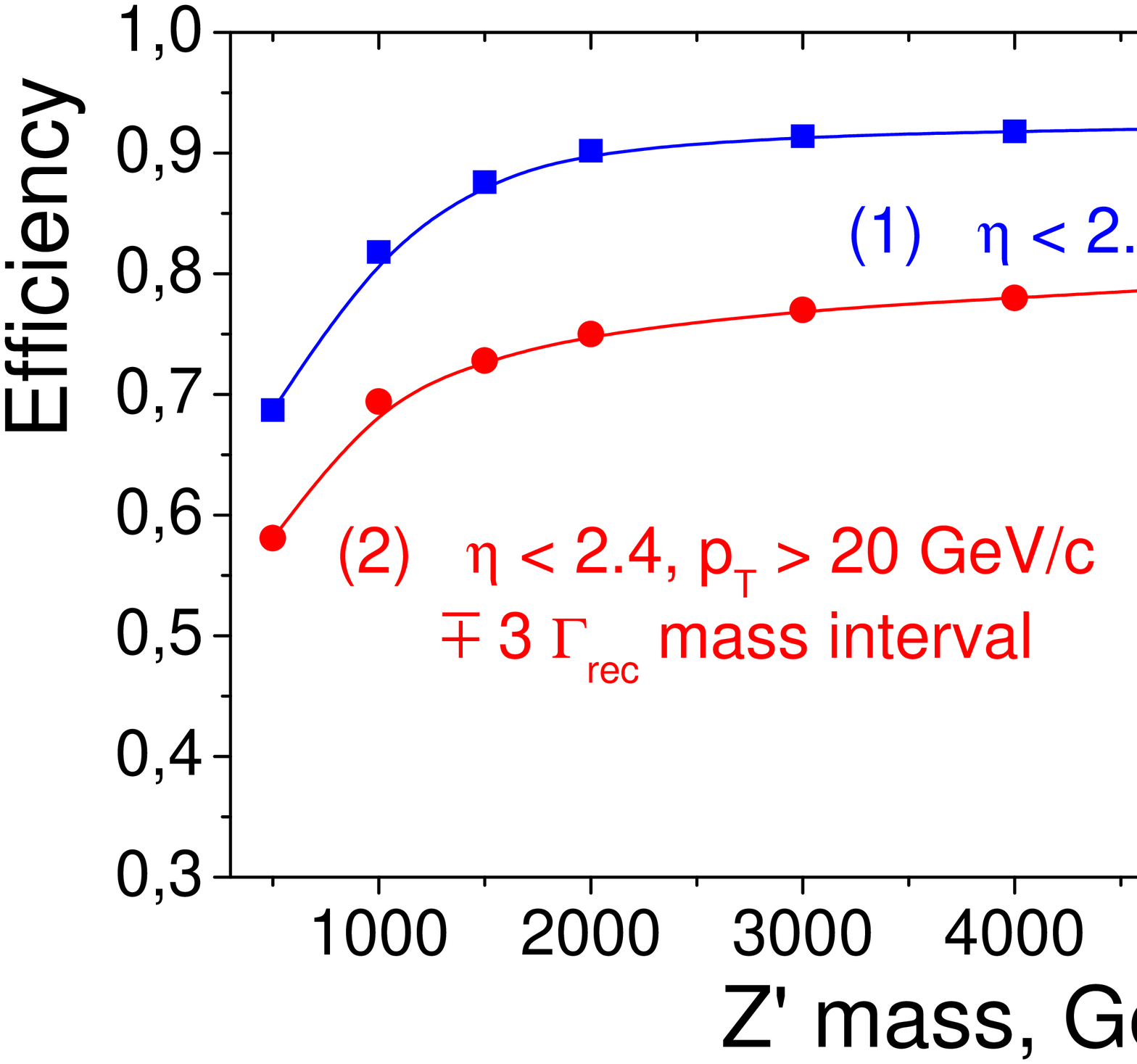}}
\resizebox{6.2cm}{!}{\includegraphics{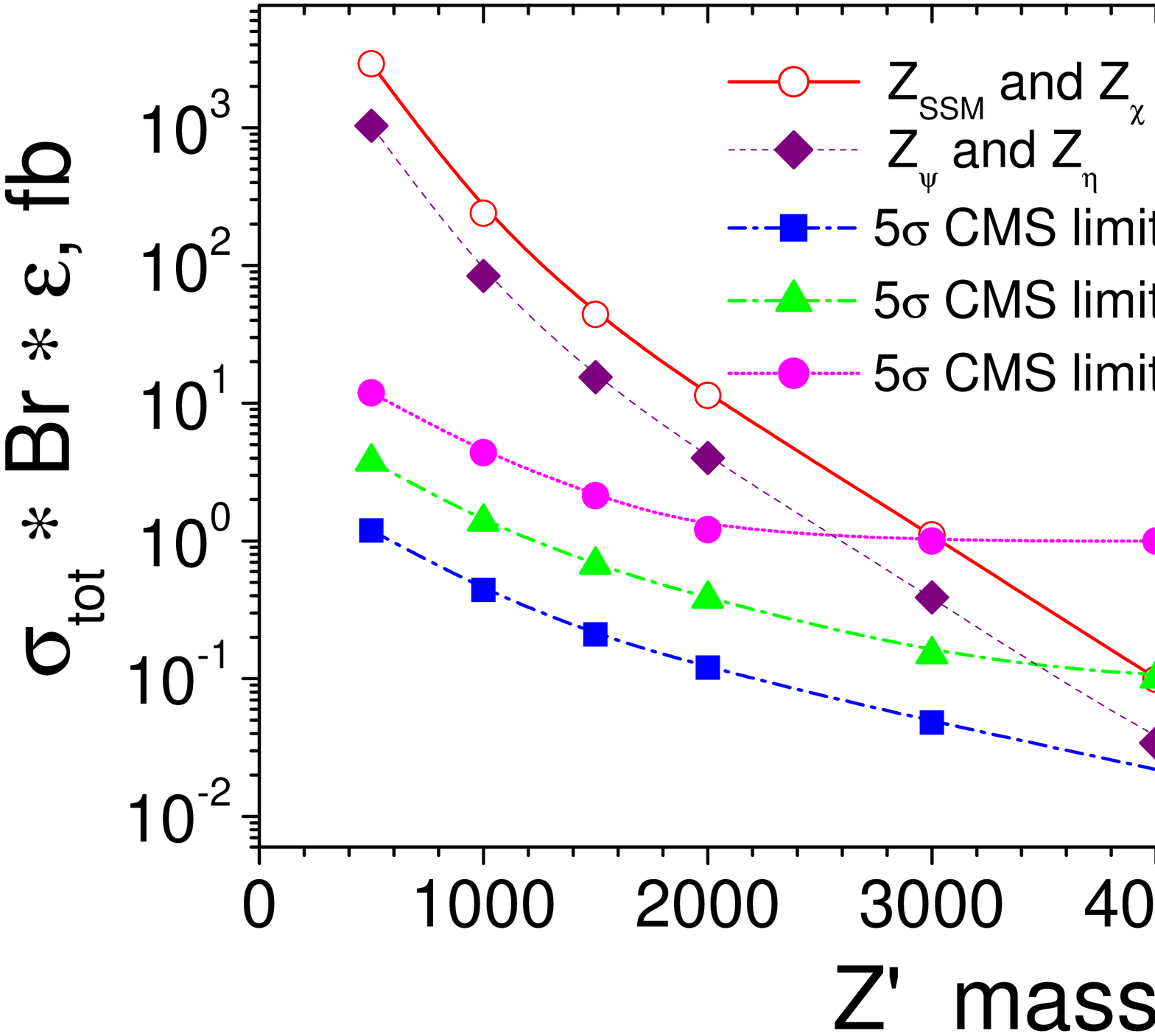}}
\vspace{-2.2cm} \caption{Efficiency of the di-muon selection as a
function the di-muon effective mass (left). The product of the
cross section and the branching ratio for the $\mbox{Z}^{\prime}$
decay into muon pairs (right). The five $\sigma$ LHC limits to
observe these states are also presented for various integrated
luminosities.} \label{fig:eff_Z}
\end{center}
\end{figure}

Event samples for seven mass values were generated with the
above-mentioned model parameters. To take into account the
detector response the parametrization of the muon momentum ($p$)
resolution, $4\%/\sqrt{p/\mbox{TeV}}$ \cite{CMS_Muon_TDR}, was
used. The di-muon is accepted when both decay muons are within
detector system covering the pseudorapidity region of $|\eta| \le
2.4$ with the efficiency presented in the Fig.~\ref{fig:eff_Z}
(left plot, upper line). No cuts are made on the isolation of
muons in the tracker and calorimeter. For each mass point, muon
pairs were selected inside the mass window defined as $\pm
3\Gamma_{obs}$ around the $\mbox{Z}^{\prime}$ mass. The
$\Gamma_{obs}$ is defined as the observed width of the resonance
state after smearing due to detector effects. In addition, the cut
$p_{\mbox{\small T}} \ge $ 20 $\mbox{GeV}/c$ was applied on each
muon. The total efficiency of dimuon selection, $\epsilon$, is
given in Fig.~\ref{fig:eff_Z} (left plot, lower line). As
Fig.~\ref{fig:crs_Z} (right plot) shows, the detector effects lead
to significant smearing (line hatched histogram) of the initial
resonance peak (open histogram), but it is still clearly visible
over Drell-Yan background (full hatched).

The product of the $\mbox{Z}^{\prime}$ production cross section
and the branching ratio is shown in Fig.~\ref{fig:eff_Z} (right
plot) for $\mbox{Z}^{\prime}_{\mbox{\small SSM}}$-,
$\mbox{Z}^{\prime}_{\chi}$-, $\mbox{Z}^{\prime}_{\psi}$-,
$\mbox{Z}^{\prime}_{\eta}$-models (the curve corresponding
$\mbox{Z}^{\prime}_{\mbox{\small LR}}$-model is not drawn). The
efficiency of di-muon pair selection, $\epsilon$, is also taken
into account.


\vspace{-0.5cm}

To estimate the $\mbox{Z}^{\prime}$ discovery limit the expected
significance of the signal, S/$\sqrt{B}$, was computed, where $S$
is the number of signal events passed through all kinematics cuts
and $B$ is the number of background events. In order to insure the
statistical significance of the signal for low background the
minimal number of signal events, $S_{\mbox{min}}$, equal to ten
has been required, i.e. $S_{\mbox{min}}$ = $\mbox{max}(5\sqrt{B},
10)$.

The discovery limits for a five $\sigma$ signal are presented in
Fig.~\ref{fig:eff_Z} (right plot) for various integrated
luminosity: 10 $\mbox{fb}^{-1}$, 100 $\mbox{fb}^{-1}$, 1000
$\mbox{fb}^{-1}$ (three bottom-up curves). As shown in this
picture, the $\mbox{Z}^{\prime}$ boson can be detected up to
masses given in Table \ref{tab:Z_limit}.

The detection of the $\mbox{Z}^{\prime}$ peak and the very precise
measurement of its mass and width do not allow the theoretical
model describing the $\mbox{Z}^{\prime}$ to be identified. To test
the helicity structure of the boson and discriminate between the
$\mbox{Z}^{\prime}$ models, the leptonic forward-backward
asymmetry can be used (see, for an example, \cite{A_FB}). The
asymmetry is defined as the ratio $A_{FB} = \frac{(F-B)}{(F +
B)}$, where $F$ and $B$ are the number of the events in the
forward and backward direction, respectively. Forward (backward)
is defined as the hemisphere with $\cos(\theta) > 0$
($\cos(\theta) < 0$), where $\theta$ is the angle between the
outgoing negative lepton and the quark $\mbox{q}$ in the $\mbox{q}
\bar{\mbox{q}}$ rest frame. Such definition assumes that the
original quark direction is known, but this is not the case for
the $\mbox{pp}$-experiment. In Ref.~\cite{AFB_LHC}, however, it
was shown that it is possible to approximate the quark direction
with the boost direction of the di-muon system with respect to the
beam axis.

\vspace{-0.8cm}
\begin{table}[ht]
\caption{The $\mbox{Z}^{\prime}$ search reach for LHC in
$\mbox{TeV}/c^2$.} \label{tab:Z_limit}
\end{table}
\vspace{-0.5cm}
\begin{table}[hbt]
\begin{center}
\hspace{0cm}
\begin{tabular}{|c|c|c|c|}
\hline
   Models               & 10 $\mbox{fb}^{-1}$ & 100 $\mbox{fb}^{-1}$ & 1000 $\mbox{fb}^{-1}$ \\
\hline
$\mbox{SSM}, \chi$      & 3.01         & 3.96          & 5.0  \\
\hline
$\psi, \eta$            & 2.54         & 3.46          & 4.48  \\
\hline
$\mbox{LR}$             & 2.96         & 4.08          & 5.35  \\
\hline
\end{tabular}
\end{center}
\end{table}

\begin{figure}[hbt]
\vspace{1.5cm}
\begin{center}
\resizebox{6.2cm}{!}{\includegraphics{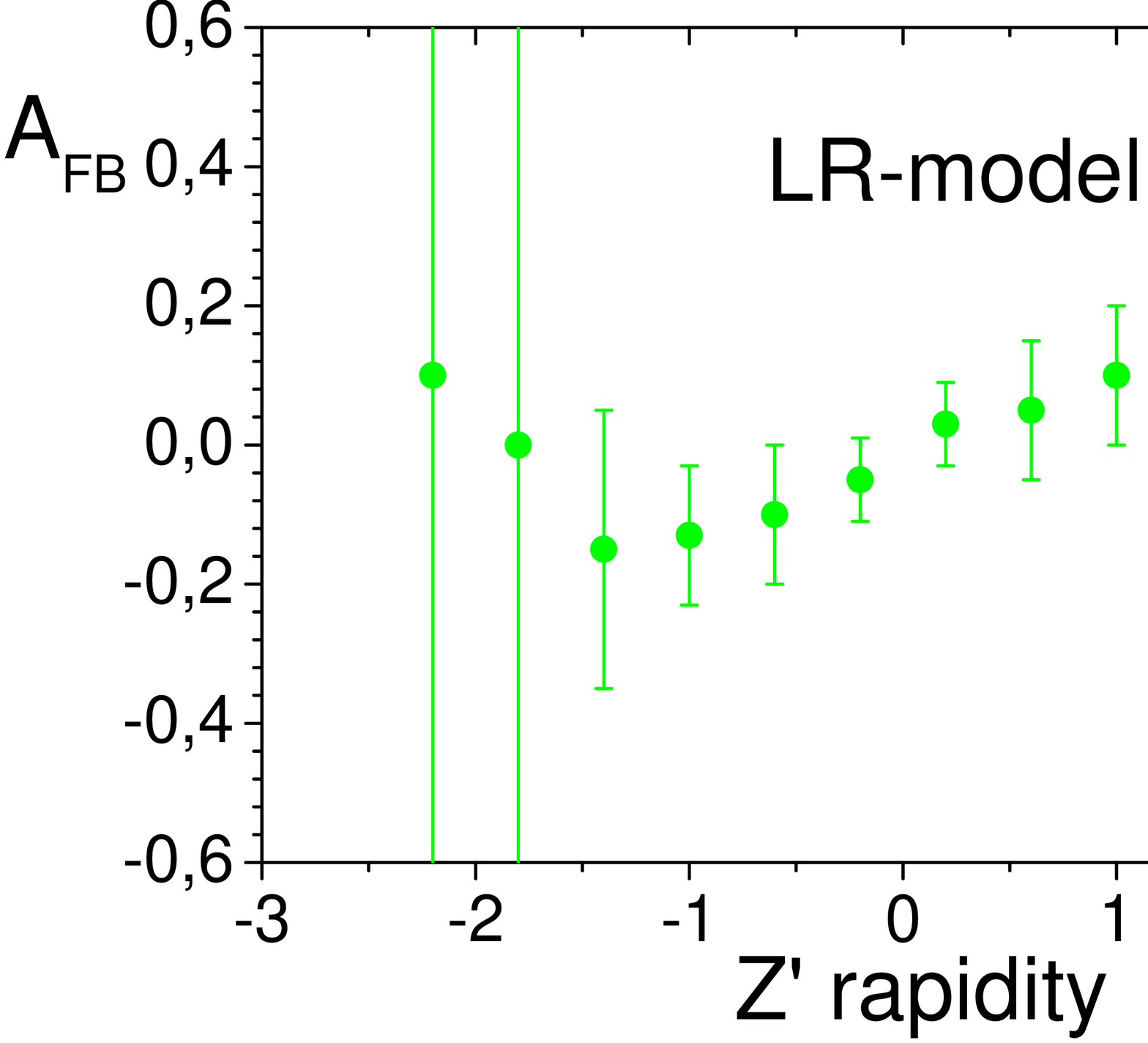}}
\resizebox{6.2cm}{!}{\includegraphics{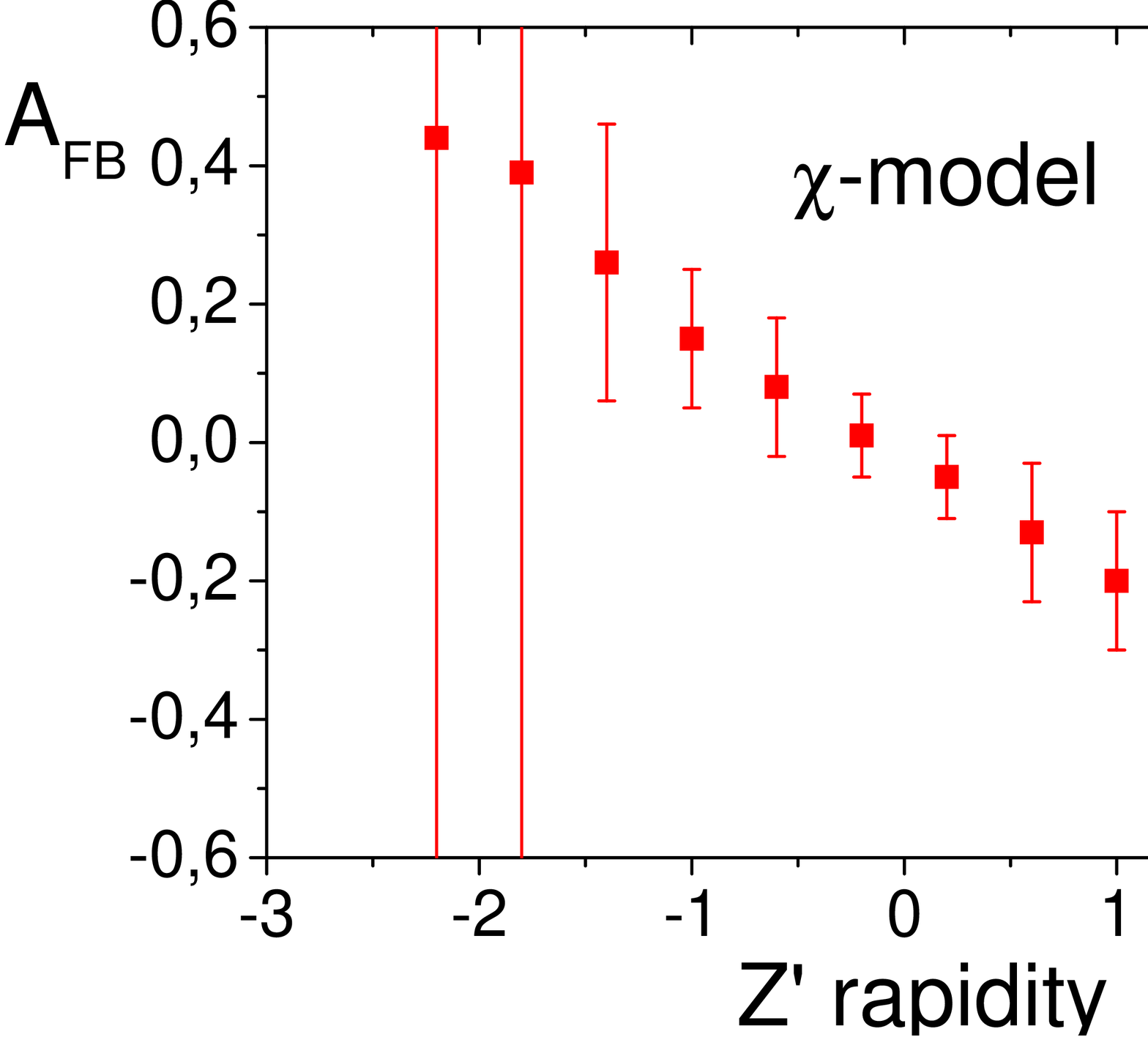}}
\vspace{-2.2cm} \caption{Muon forward-back asymmetry as function
of the $\mbox{Z}^{\prime}$ rapidity, $M_{\mbox{Z}^{\prime}}$ = 2.0
$\mbox{TeV}/c^2$.} \label{fig:Zprime_asymm}
\end{center}
\end{figure}

One of the features of the asymmetry, $A_{FB}$, is the distinctive
rapidity-dependence for different $\mbox{Z}^{\prime}$ models. Such
dependence is shown in Fig.~\ref{fig:Zprime_asymm} for the
$\mbox{Z}^{\prime}_{\mbox{\small LR}}$ (left picture) and the
$\mbox{Z}^{\prime}_{\chi}$ (right picture) under assumption of a
mass $M_{\mbox{\small Z}^{\prime}}$ = 2.0 $\mbox{TeV}/c^2$, for
100 $\mbox{fb}^{-1}$ integrated luminosity.

\vspace{-0.5cm}

\section{RS1 graviton}

\subsection{Signal and background simulations}

The ability to test experimentally the RS1 scenario predictions
depends on the  model parameter $c=k/M_{Pl}$ which controls the
coupling of the graviton to the ordinary particles and the width
of the resonance $\Gamma \sim \rho m_0 c^2,$ where the constant
$\rho$ is determined by the number of open decay channels. The
theoretical limitations give the allowed range for the coupling
constant $0.01 \le c \le 0.1$.

The graviton resonances can be produced {\it virtually} via
quark-antiquark annihilation $\mbox{q} \bar{\mbox{q}} \to
\mbox{G}_{\mbox{\small KK}}$ as well as gluon-gluon fusion
$\mbox{gg} \to \mbox{G}_{\mbox{\small KK}}$. The first of these
processes is identical to the Standard Model $s$-channel exchange
of an intermediate $\gamma^{\star}$ or $Z$ vector boson, while the
second one has no SM analogue. Other partonic sub-processes are
also possible, $\mbox{gg} \to \mbox{g G}_{\mbox{\small KK}}$,
$\mbox{q} \bar{\mbox{q}} \to \mbox{g} \mbox{G}_{\mbox{\small
KK}}$, $\mbox{gq} \to \mbox{q} \mbox{G}_{\mbox{\small KK}}$, which
form a {\it real} graviton via the $t$-channel exchange (graviton
emission).

To simulate both real and virtual graviton production in the
proton-proton collisions at 14 TeV center-of-mass energy, PYTHIA
6.217 was used in which the RS1 scenario was implemented with
CTEQ5L parton distribution functions.

The graviton production cross section for all five possible
diagrams is presented in Table~\ref{tab:RS1_crs}. Here, two
opposite possibilities for model parameter $c$ were considered.
For the first case, which is the most difficult case for
experimental detection, the constant $c=0.01$ was used (the number
in the brackets), and the second, the most optimistic, scenario
when $c$ is equal to 0.1. The majority (at least 50 $\div$ 60 \%
depending on the mass) of the gravitons is produced in processes
of gluon-gluon fusion with real graviton emission, whereas the
virtual graviton production adds up to 15 \% only of the total
cross-section.

\vspace{-0.6cm}
\begin{table}[h,pt]
\caption{Cross sections of $\mbox{G}_{\mbox{\small KK}}$
production in fb. The CTEQ5L parton distributions and $K$-factor=1
have been used.} \label{tab:RS1_crs}
\end{table}
\vspace{-0.2cm}
\begin{table}[h,pt]
\begin{center}
\hspace{0cm}
\begin{tabular}{|c|c|c|c|}
\hline
Mass, $\mbox{TeV}/c^2$                                              & $1.0$       &  $1.5$     & $3.0$ \\
\hline
$\mbox{q} \bar{\mbox{q}} \to \mbox{G}_{\mbox{\small KK}}$           & 129 (1.34)  & 23 (0.24)  & 0.633 (0.006)  \\
\hline
$\mbox{gg} \to \mbox{G}_{\mbox{\small KK}}$                         & 567 (5.33)  & 62 (0.53)  & 0.94 (0.004)  \\
\hline
$\mbox{q} \bar{\mbox{q}} \to \mbox{g} \mbox{G}_{\mbox{\small KK}}$  & 345 (3.29)  & 65 (0.64)  & 1.84 (0.017)  \\
\hline
$\mbox{qg} \to \mbox{q} \mbox{G}_{\mbox{\small KK}}$                & 599 (5.78)  & 72 (0.64)  & 1.05 (0.007)  \\
\hline
$\mbox{gg} \to \mbox{g} \mbox{G}_{\mbox{\small KK}}$                & 3350 (31.5) & 368 (3.32) & 4.98 (0.028)  \\
\hline
Total                                                               & 4990 (47.2) & 590 (5.38) & 9.45 (0.062) \\
\hline
\end{tabular}
\end{center}
\end{table}
\vspace{0.5cm}

The Standard Model background for this channel is the same as for
the $\mbox{Z}^{\prime}$ case.

\subsection{Detection of RS1-graviton resonance and discovery limits}

\begin{figure}[ht]
\vspace{1.5cm}
\begin{center}
\resizebox{6.2cm}{!}{\includegraphics{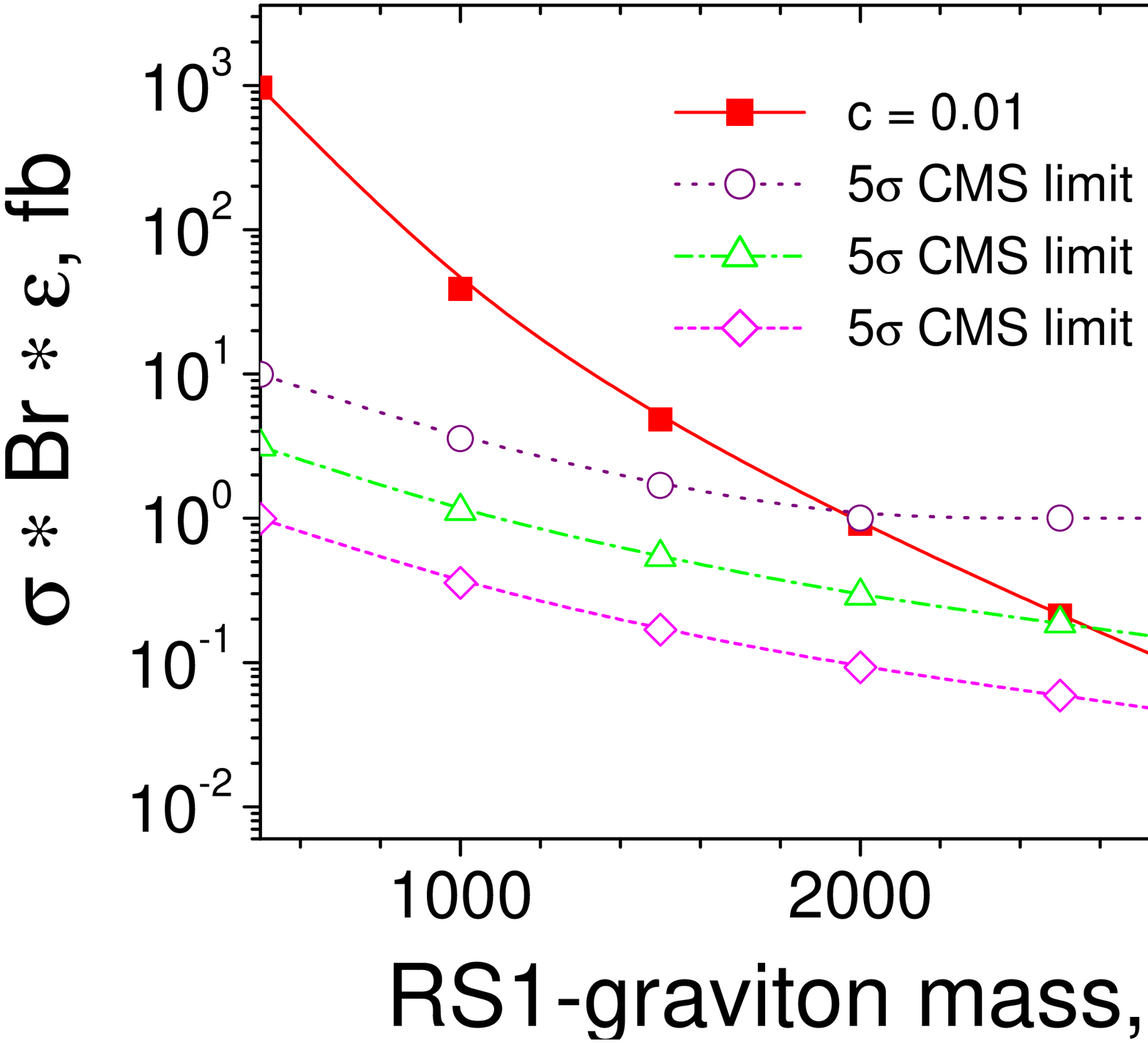}}
\resizebox{6.2cm}{!}{\includegraphics{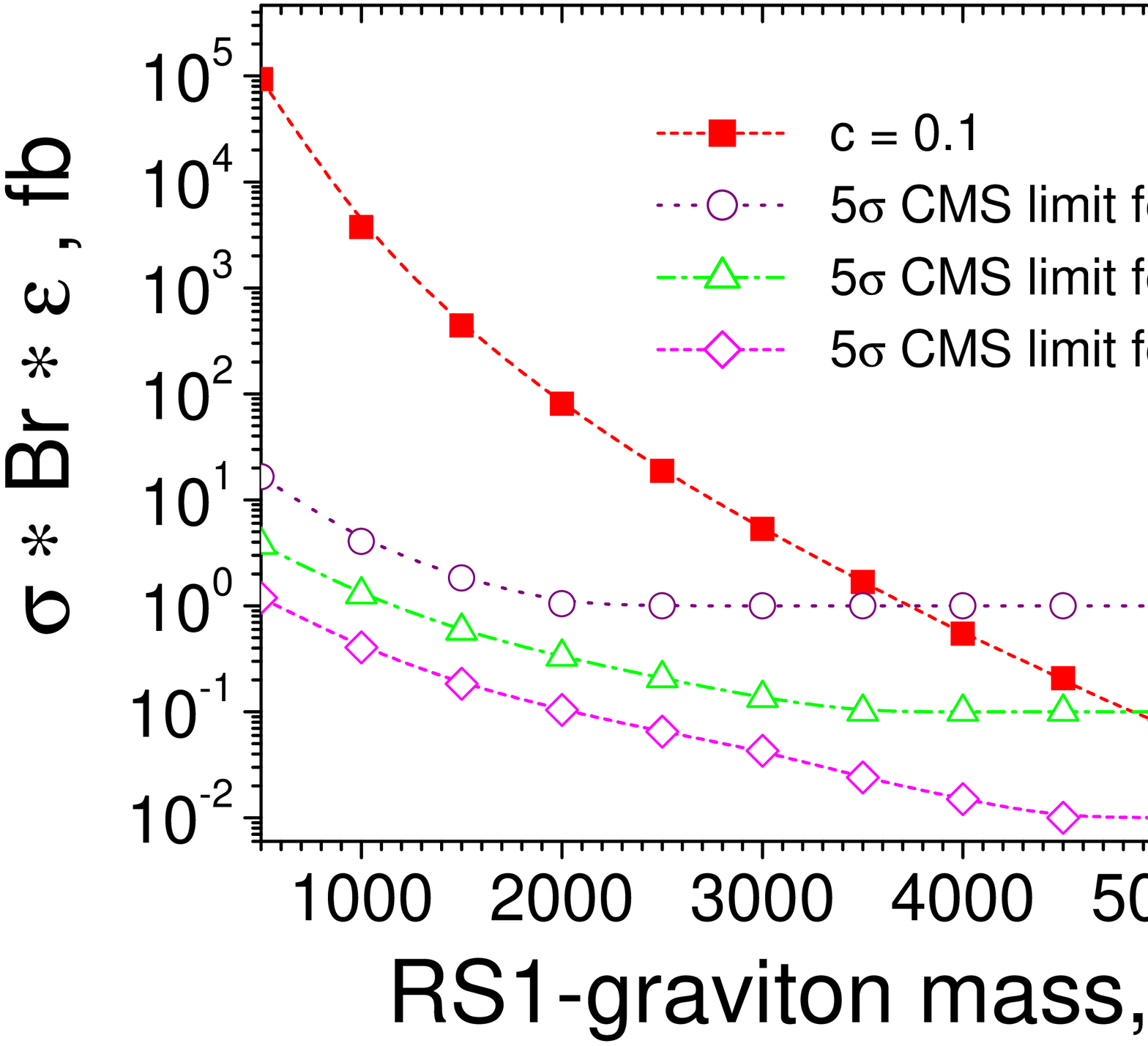}}
\vspace{-2.2cm} \caption{Cross section of the
$\mbox{G}_{\mbox{\small KK}}$ production {\it vs} their mass for
$c$ = 0.01 (left) and $c$ = 0.1 (right). The five $\sigma$ limits
to observe these states are also presented for various integrated
luminosities.} \label{fig:RS1_sign}
\end{center}
\end{figure}

The simulation of the detector response for the graviton decay
into muon pair is similar to the $\mbox{Z}^{\prime}$ case
(Sect.~2.2). To estimate the discovery limit for the RS1 graviton
the same procedure as for the $\mbox{Z}^{\prime}$ case was
applied. The cross section of $\mbox{G}_{\mbox{\small KK}}$
production and the corresponding cross section limits to observe a
five $\sigma$ signal, for various integrated luminosity, are
presented in Fig.~\ref{fig:RS1_sign}. As shown in the figure, the
LHC can test the RS1 scenario in the whole range of the model
parameter ($c$ = 0.01) up to a mass of 2.0 $\mbox{TeV}/c^2$ even
with a low luminosity of 10 $\mbox{fb}^{-1}$. In the more
favourable case with $c$ = 0.1 the accessible mass region is
extended up to 3.7 $\mbox{TeV}/c^2$. With 100 $\mbox{fb}^{-1}$ the
reach increases to 2.6 $\mbox{TeV}/c^2$ for $c$ = 0.01 and 4.8
$\mbox{TeV}/c^2$ for $c$ = 0.01.

The results of the combined analysis in the RS1 scenario
\cite{Restr_coupl} show that the value of the dimensionless
coupling constant $c$ and the corresponding value of the graviton
mass are strongly restricted due to the experimental Tevatron data
and theoretical constraints to assure the model hierarchy
($\Lambda_{\pi} <$ 10 TeV). The limitations lead, in particular,
to the conclusion that the constant $c$ can not be less than
0.027, for the graviton mass of one $\mbox{TeV}/c^2$, and less
than 0.1, for the mass of 3.7 $\mbox{TeV}/c^2$. The direct
comparison of these results with the data of the
Fig.~\ref{fig:RS1_sign} shows that the whole space of the
RS1-model parameters is accessible at luminosity of 100
$\mbox{fb}^{-1}$, and the RS1 graviton can be discovered with the
five $\sigma$ significance. These conclusions, however, are not
definitive, since the initial theoretical constraints are very
arbitrary.

\vspace{-0.5cm}

\begin{figure}[ht]
\vspace{0.5cm}
\begin{center}
\resizebox{6cm}{!}{\includegraphics{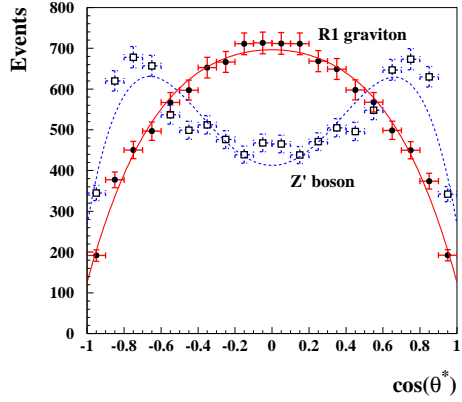}} \caption{Angular
distributions $\cos(\theta^{\star})$ of the muons from the
$\mbox{G}_{\mbox{\small KK}}$ (solid marker) and the
$\mbox{Z}^{\prime}$ (open box) decays.} \label{fig:angle}
\end{center}
\end{figure}

Under the assumption that the new resonance state is observed at
LHC, the nature of this object should be understood (in principle,
it can come from the extended gauge sector as well as from any
version of extra dimensions). The major difference between the
$\mbox{Z}^{\prime}$ and the RS1 graviton should appear in the
$\cos(\theta^{\star})$ distribution (where $\theta^{\star}$ is the
polar angle of the muons in the center-of-mass system of the
di-muon pair) which is strongly spin dependent. Certainly, these
distributions will be distorted by acceptance cuts, especially in
the region of large angles, and the expected theoretical
predictions will differ from experimental ones. Nevertheless
Fig.~\ref{fig:angle} shows a distinct difference between the
spin-1 ($\mbox{Z}^{\prime}$) and the spin-2 (RS1 graviton) curves
obtained for muons after all cuts. To ease the comparison, these
plots were obtained for resonance states with mass of 1.5
$\mbox{TeV}/c^2$, and normalized to 6000 events that correspond to
an approximate integrated luminosity of 10 $\mbox{fb}^{-1}$ for
graviton production with $c$ = 0.1 and 100 $\mbox{fb}^{-1}$ for
$\mbox{Z}^{\prime}$ boson.

\vspace{0.5cm}

\section{Summary}

This work presents the discovery potential of $\mbox{Z}^{\prime}$
gauge bosons as well as RS1 gravitons in the muon channel at the
LHC experiments. The estimated discovery limit for
$\mbox{Z}^{\prime}$ is about 3.5 $\div$ 4.0 $\mbox{TeV}/c^2$,
depending on the couplings, for 100 $\mbox{fb}^{-1}$ integrated
luminosity. At the same luminosity, zero KK-modes of the RS1
graviton state can be observed up to 2.6 $\mbox{TeV}/c^2$ and 4.8
$\mbox{TeV}/c^2$ for $c$ = 0.01 and 0.1, respectively. The angular
distribution of the muons in the final state can be used to
distinguish the spin-1 and the spin-2 hypotheses, at least in the
mass region up to 1.5 $\mbox{TeV}/c^2$. The different
$\mbox{Z}^{\prime}$ models can be distinguished (up to mass of two
$\mbox{TeV}/c^2$) because of the different the leptonic
forward-backward asymmetry.

Further detailed studies are required on the theoretical side in
order to identify other possible physics observables. This would
be helpful to better understand the LHC discovery conditions and
limits.

\bigskip

{\small This work was supported in part by INTAS under project
YSF-2001/1-218. We would like to thank E.~Boos, D.~Denegri,
M.~Dubinin, A.~Lanyov, L.~Pape, M.~Spira, and S.~Valuev for
enlightening and helpful discussions. We are also grateful to
D.~Acosta, V.~Gavrilov, and P.~Sphicas for their censorious and
actual remarks and friendly critics concerning the presented
talk.}

\end{document}